\documentclass[12pt]{article}
\usepackage[dvips]{color}
\usepackage{epsfig}
\usepackage{amsmath}
\usepackage{graphicx}
\usepackage{cite}
\usepackage{subcaption}

\begin{document}

\title{\bf The extended uncertainty principle effects on the phase transitions of Reissner-Nordstr\"{o}m and Schwarzschild black holes}
\author{{ \"{O}zg\"{u}r \"{O}kc\"{u} \thanks{Email: ozgur.okcu@ogr.iu.edu.tr}\hspace{1mm},
Ekrem Aydiner \thanks{Email: ekrem.aydiner@istanbul.edu.tr}} \\
{\small {\em Department of
		Physics, Faculty of Science, Istanbul University,
		}}\\{\small {\em Istanbul, 34134, Turkey}} }
\maketitle

\begin{abstract}
In this paper, we investigate the phase transitions of Reissner-Nordstr\"{o}m
			(RN) and Schwarzschild black holes for the extended uncertainty principle
			(EUP) framework. Considering temperature $T$, charge $Q$ and electric potential
			$\Phi $ as the state parameters, we show the van der Waals (vdW) like phase
			transition of RN black hole in $Q-\Phi $ diagrams and find the critical
			points depending on EUP parameter $\alpha $. Furthermore, we find Hawking-Page
			like phase transition for Schwarzschild black hole. The results imply that
			the black holes in asymptotically flat space have the similar phase structure
			with the black holes in anti-de Sitter (AdS) space.\\

\end{abstract}

\section{Introduction}

\label{sec1}
	
	The black hole thermodynamics is one of the most interesting subject in
	the theoretical physics. Black holes have been considered as thermodynamic
	systems since the breakthrough studies of Bekenstein and Hawking
	\cite{Bekenstein1972,Bekenstein1973,Bardeen1973,Hawking1974,Bekenstein1974,Hawking1975}.
	The black hole thermodynamics may play a vital role for understanding the
	quantum gravity theories. It may also provide the fundamental relations
	between general relativity, thermodynamics and quantum mechanics. Besides,
	considering black holes with a temperature and an entropy reveals many
	interesting similarities with the conventional thermodynamics systems.
	Specifically, similar phase transitions and critical phenomena can be found
	for the black holes. The studies in this direction were pioneered by Davies
	who considered the phase transition of Kerr-Newmann black holes
	\cite{Davies1977}. Hawking and Page later showed a phase transition between
	Schwarzschild-AdS black hole and thermal AdS space \cite{Hawking1983}.
	Once they paved the way to phase transition of black holes, many papers
	devoted to this direction have been widely studied in the literature and
	new phase structures have been discovered
	\cite{Chamblin1999a,Chamblin1999b,Banerjee2010,Banerjee2010b,Banerjee2011,Banerjee2012,Lala2012,Banerjee2011b,Niu2012,Ma2016,Wei2013,Tsai2012,Deng2017,Deng2021,Kubiznak2012,Gunasekaran2012,Zhao2013,Cai2013,Mo2014,Mo2014b,Li2014,Hendi2017,Belhaj2015,Caceres2015,Wei2016b,Bhatt2017,Altamirano2014,Frassino2014,Hennigar2015,Wei2014}.
	In Refs. \cite{Chamblin1999a,Chamblin1999b}, Chamblin et al. studied the
	thermodynamics of charged AdS black hole and showed a vdW like phase transition.
	In Refs.
	\cite{Banerjee2010,Banerjee2010b,Banerjee2011,Banerjee2011b,Lala2012,Banerjee2012},
	the authors proposed the novel Ehrenfest equations by considering the analogies
	$V\leftrightarrow Q$, $P\leftrightarrow \Phi $ or
	$V\leftrightarrow J$, $P\leftrightarrow \Omega $. Furthermore,
	$Q-\Phi $ and $J-\Omega $ criticalities revealed more analogies between
	the AdS black holes and vdW fluids
	\cite{Niu2012,Ma2016,Wei2013,Tsai2012,Deng2017,Deng2021}.
	
	Moreover, these analogies become more exact in the extended phase space
	where the cosmological constant and its conjugate quantities are identified as (we use the units $\hbar =G_{N}=c=k_{B}=L_{Pl}=1$ throughout the paper.)
	%
	\begin{equation}
		\label{pressure}
		P=-\frac{\Lambda}{8\pi}\,,
	\end{equation}
	%
	\begin{equation}
		\label{thermoVolume}
		V=\left (\frac{\partial M}{\partial P}\right )_{S,Q,J}\,,
	\end{equation}
	pressure and thermodynamic volume, respectively. Based on this fact, Kubiznak and Mann showed the $P-V$ criticality of the charged AdS black holes \cite{Kubiznak2012}. They obtained the critical points and showed that the critical exponents of the charged black holes are identical to the
	exponents of vdW fluids. Then, the same phase transition was studied for
	different AdS black holes
	\cite{Gunasekaran2012,Zhao2013,Cai2013,Mo2014,Mo2014b,Li2014,Hendi2017,Belhaj2015,Caceres2015,Wei2016b,Bhatt2017}.
	Also the existence of triple points and reentrant phase transitions for
	AdS black holes were reported in Refs.
	\cite{Altamirano2014,Frassino2014,Hennigar2015,Wei2014} (for a more comprehensive
	review, please see Ref.~\cite{Kubiznak2017}).
	
	On the other hand, black hole thermodynamics and phase transitions can
	be considered beyond the semiclassical approximation. Considering the first
	order correction to entropy and temperature, authors studied the
	$\Omega -J$ criticality for Myers-Perry black holes in Ref.~\cite{Poshteh2013}. In Ref.~\cite{Czinner2016}, Schwarzschild black entropy
	was considered in Tsallis-Renyi approach. The authors found a Hawking-Page
	like phase transition for Schwarzschild black hole in asymptotically flat
	spacetime. vdW phase transition of charged black hole in flat space was
	studied for Renyi statistics in Ref.~\cite{Promsiri2020}.
	
	The modifications of Heisenberg uncertainty principle, i.e., the generalized
	uncertainty principle (GUP) and EUP, also have some interesting results
	on the black hole thermodynamics and phase transitions. Various GUP and
	EUP models were proposed in the literature
	\cite{Maggiore1993,Scardigli1999,Kempf1995,Bambi2008,Nozari2012,Filho2016,Chung2018,Lake2019,Chung2019,Dabrowski2019,Lake2020,Lake2021,Mureika2019}.
	Since GUP plays a remarkable role in the Planck scale physics, the modifications
	of black hole thermodynamics may not be unavoidable. Therefore, various
	GUP corrections to black hole thermodynamics were investigated in the literature
	\cite{Adler2001,Medved2004,Nozari2008,Arraut2009,Banerjee2010x,Nozari2012b,Feng2016,Xiang2009,Scardigli2020a,Hassanabadi2021,Lutfuoglu2021,Sun2018,Ma2018,Okcu2020}.
	Furthermore, richer and more complicated phase structures can be considered
	in the GUP framework. In Refs. \cite{Sun2018,Ma2018}, Ma et al. studied
	GUP effects on the phase transitions of charged AdS black hole. They found
	reentrant phase transition similar to higher dimensional Kerr-AdS black
	hole case. In Ref.~\cite{Okcu2020}, we considered the GUP for vdW black
	hole. We found that $P-V$ criticality is physically acceptable for the
	GUP case.
	
	On the other hand, EUP may affect the black hole thermodynamics at large
	scales since it introduces the maximum position uncertainty notion
	\cite{Bolen2005,Han2008,MoradpourEUP2019,HassanabadiEUP2019,ChungEUP2019,HamilEUP2021,HamilEUP2021b,ChenEUP2019}.
	Based on the fact that EUP effects are not negligible at the large distance
	scale, Bolen and Cavaglia showed that Schwarzschild-(A)dS black hole temperature
	can be derived from the EUP in Ref.~\cite{Bolen2005}.\footnote{At this point,
		we bring the Ref.~\cite{Scardigli2020} to the reader's attention. The motivation
		of Ref.~\cite{Bolen2005} was criticized in Ref.~\cite{Scardigli2020}. In
		order to obtain the temperature of Schwarzschild-(A)dS black hole, Bolen
		and Cavaglia chose the EUP parameter identified with (A)dS radius. In Ref.~\cite{Scardigli2020}, Scardigli showed that the temperature of (A)dS black
		holes can heuristically be derived from Heisenberg uncertainty principle
		without any adjustable numerical constant. There is no need the EUP to
		compute (A)dS black hole temperature. Therefore, we may assume that there
		is not a necessary link between (A)dS black hole temperature and EUP.}
	In Ref.~\cite{Han2008}, modified dispersion relations and EUP effects were
	investigated for (A)dS black holes. In Ref.~\cite{MoradpourEUP2019}, the
	authors employed the EUP to obtain the Renyi entropy for the black holes.
	Recently, the black hole thermodynamics has been considered for the various
	EUP models
	\cite{HassanabadiEUP2019,ChungEUP2019,HamilEUP2021,HamilEUP2021b,ChenEUP2019}.
	
	Since (A)dS black hole thermodynamics may be related to EUP
	\cite{Bolen2005,Han2008}, EUP corrected black holes in flat spacetime may
	have the same phase structures similar to AdS case. Therefore, we would
	like to investigate the phase transition properties of Schwarzschild and
	RN black holes for the EUP case.
	
	The paper is organized as follows: In Sect.~\ref{sec2}, we obtain the EUP modified
	temperature, entropy and heat capacity of RN black holes. In Sect.~\ref{sec3}, we
	investigate vdW like phase transition in $Q-\Phi $ diagrams. We also study
	the behaviours of heat capacity and Gibbs free energy for the phase transition.
	In Sect.~\ref{sec4}, we show a Hawking-Page like phase transition for Schwarzschild
	black hole. In Sec.~\ref{sec5}, we finally discuss our results.

\section{RN black holes thermodynamics with EUP correction}
 
\label{sec2}
	
	RN black hole solution in four dimensional spacetime is given by
	\cite{Altamirano2014}
	%
	\begin{equation}
		\label{metric1}
		ds^{2}=-f(r)dt^{2}+\frac{dr^{2}}{f(r)}+r^{2}d\theta ^{2}+r^{2}\sin ^{2}
		\theta d\phi ^{2},
	\end{equation}
	where
	%
	\begin{equation}
		\label{metrik2}
		f(r)=1-\frac{2M}{r}+\frac{Q^{2}}{r^{2}}.
	\end{equation}
	From $f(r_{h})=0$, we obtain the black hole mass
	%
	\begin{equation}
		\label{mass}
		M=\frac{r_{h}}{2}\left (1+\frac{Q^{2}}{r_{h}^{2}}\right ),
	\end{equation}
	where $r_{h}$ is the event horizon of black hole.
	
	In order to obtain the EUP correction, we consider the analysis of Xiang
	and Wen \cite{Xiang2009}. Now, let us consider the first law of black hole
	thermodynamics
	%
	\begin{equation}
		\label{firstLaw}
		dM=TdS+\Phi dQ.
	\end{equation}
	Since entropy is generally function of horizon area A, temperature of black
	hole can be written as
	%
	\begin{equation}
		\label{tempBlackHoles}
		T=\left (\frac{\partial M}{\partial S}\right )_{Q}=\frac{dA}{dS}
		\times \left (\frac{\partial M}{\partial A}\right )_{Q}=\frac{dA}{dS}
		\times \frac{\kappa}{8\pi},
	\end{equation}
	where $\kappa $ is the surface gravity of black hole and is defined by
	%
	\begin{equation}
		\label{surfaceGravity}
		\kappa =\frac{f'(r_{h})}{2}=\frac{r_{h}^{2}-Q^{2}}{2r_{h}^{3}}.
	\end{equation}
	When a particle is absorbed by a black hole, the smallest increase of area
	is given by \cite{Bekenstein1973}
	%
	\begin{equation}
		\label{areaInC0}
		\Delta A \sim bm
	\end{equation}
	where $b$ and $m$ correspond to particle's size and mass, respectively.
	In order to define the Eq.(\ref{areaInC0}) in a more familiar form, we
	consider the limitations on $b$ and $m$. In quantum mechanics, we define
	a particle by a wave packet. The width of a wave packet is characterized
	by the size of particle $b$, i.e., $b\sim \Delta x$. On the other hand,
	the momentum uncertainty is not greater than the particle mass $m$, i.e.,
	$m\leq \Delta p$. Therefore, we can write
	%
	\begin{equation}
		\label{areaInc}
		\Delta A\sim bm\geq \Delta x \Delta p.
	\end{equation}
	We consider the simplest EUP \cite{Mureika2019},
	%
	\begin{equation}
		\Delta x\Delta p\geq 1+\alpha \Delta x^{2},
	\end{equation}
	where the EUP parameter $\alpha $ is defined by
	$\alpha =\alpha _{0}/L_{*}^{2}$, $\alpha _{0}$ is the dimensionless positive
	constant and $L_{*}$ is the new fundamental length scale. Using the EUP
	and taking $\Delta x\sim 2r_{h}$, one can get
	%
	\begin{equation}
		\label{AreaEUP}
		\Delta A\geq \gamma \left (1+4\alpha r_{h}^{2}\right ),
	\end{equation}
	where $\gamma $ is a calibration factor which is determined from standard
	results. The minimum increase of entropy is given by
	$(\Delta S)_{min}=\ln 2$. Therefore, the EUP modified entropy-area relation
	is given by
	%
	\begin{equation}
		\label{entAreaRel}
		\frac{dA}{dS}\simeq \frac{\Delta A_{min}}{\Delta S_{min}}=
		\frac{\gamma \left (1+4\alpha r_{h}^{2}\right )}{\ln 2}.
	\end{equation}
	In the limit $\alpha \rightarrow 0$, the Eq. (\ref{entAreaRel}) must give
	$dA/dS=4$. So we find $\gamma =4\ln 2$. Using Eqs. (\ref{tempBlackHoles}),
	(\ref{surfaceGravity}) and (\ref{entAreaRel}), the EUP corrected black
	hole temperature is obtained as
	%
	\begin{equation}
		\label{EUPTemperature}
		T=
		\frac{\left (r_{h}^{2}-Q^{2}\right )\left (1+4\alpha r_{h}^{2}\right )}{4\pi r_{h}^{3}}.
	\end{equation}
	The EUP corrected entropy and heat capacity are given by
	%
	\begin{equation}
		\label{entropy}
		S=\int \frac{dM}{T}=\frac{\pi}{4\alpha}\ln \left (1+4\alpha r_{h}^{2}
		\right ),
	\end{equation}
	%
	\begin{equation}
		\label{heatCapa}
		C_{Q}=T\left (\frac{\partial S}{\partial T}\right )_{Q}=
		\frac{2\pi r_{h}^{2}\left (r_{h}^{2}-Q^{2}\right )}{3Q^{2}-r_{h}^{2}+4\alpha r_{h}^{2}\left (Q^{2}+r_{h}^{2}\right )},
	\end{equation}
	respectively. Finally, the electric potential is given by
	%
	\begin{equation}
		\label{elecPot}
		\Phi =\left (\frac{\partial M}{\partial Q}\right )_{S}=
		\frac{Q}{r_{h}}.
	\end{equation}

\section{$Q-\Phi$ criticality of RN black holes}
\label{sec3}

\begin{figure}
        \centering
		\includegraphics[width=13cm]{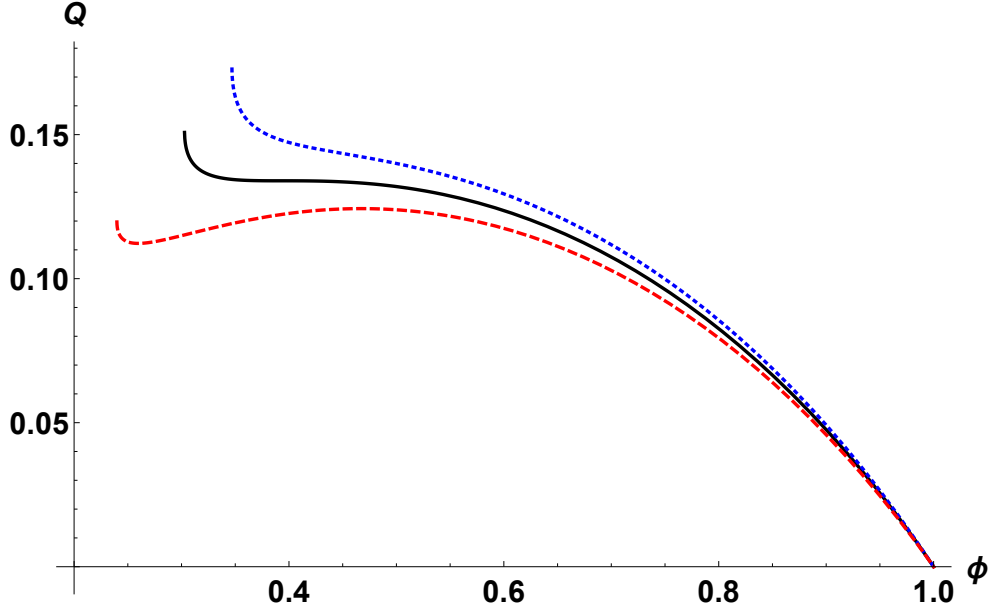}
		\caption{$Q-\Phi $ diagram of RN black hole for the EUP case.
			Blue dotted isotherm corresponds to $T<T_{c}$. The black solid isotherm is for
			the critical temperature. Red dashed isotherm corresponds to $T>T_{c}$. We take
			$\alpha =1$.}
		\label{f1}
	\end{figure}
	
	Using Eqs. (\ref{EUPTemperature}) and (\ref{elecPot}), we obtain the equation
	of state $Q=Q(T,\Phi )$ as
	%
	\begin{equation}
		\label{equationOfState}
		Q=\Phi
		\frac{\pi T-\sqrt{\pi ^{2}T^{2}-\alpha \left (\Phi ^{2}-1\right )^{2}}}{2\alpha \left (1-\Phi ^{2}\right )}.
	\end{equation}
	The critical points can be obtained from the derivatives of Eq. (\ref{equationOfState})
	with respect to electric potential, i.e.,
	%
	\begin{equation}
		\label{cP}
		\left (\frac{\partial Q}{\partial \Phi}\right )_{T}=0,\qquad \left (
		\frac{\partial ^{2}Q}{\partial \Phi ^{2}}\right )_{T}=0.
	\end{equation}
	From Eqs. (\ref{equationOfState}) and (\ref{cP}), we find
	%
	\begin{equation}
		\label{Tc}
		T_{c}=\frac{4\sqrt{\left (2\sqrt{3}-3\right )\alpha}}{3\pi},
	\end{equation}
	%
	\begin{equation}
		\label{PHIc}
		\Phi _{c}=\sqrt{\frac{2\sqrt{3}-3}{3}},
	\end{equation}
	%
	\begin{equation}
		\label{Qc}
		Q_{c}=\frac{2-\sqrt{3}}{2\sqrt{\alpha}}.
	\end{equation}
	At this point, we give some comments on the critical points. The critical
	temperature and charge depend on the EUP parameter $\alpha $, but the critical
	electric potential does not depend on $\alpha $. A similar case was also
	found for $(n+1)$-dimensional RN-AdS black holes in Ref.~\cite{Ma2016} where the authors find the critical temperature and charge
	depend on $\Lambda $ while the critical electric potential does not depend
	on $\Lambda $.
	
\begin{figure}
	\centering
		\includegraphics[width=8cm]{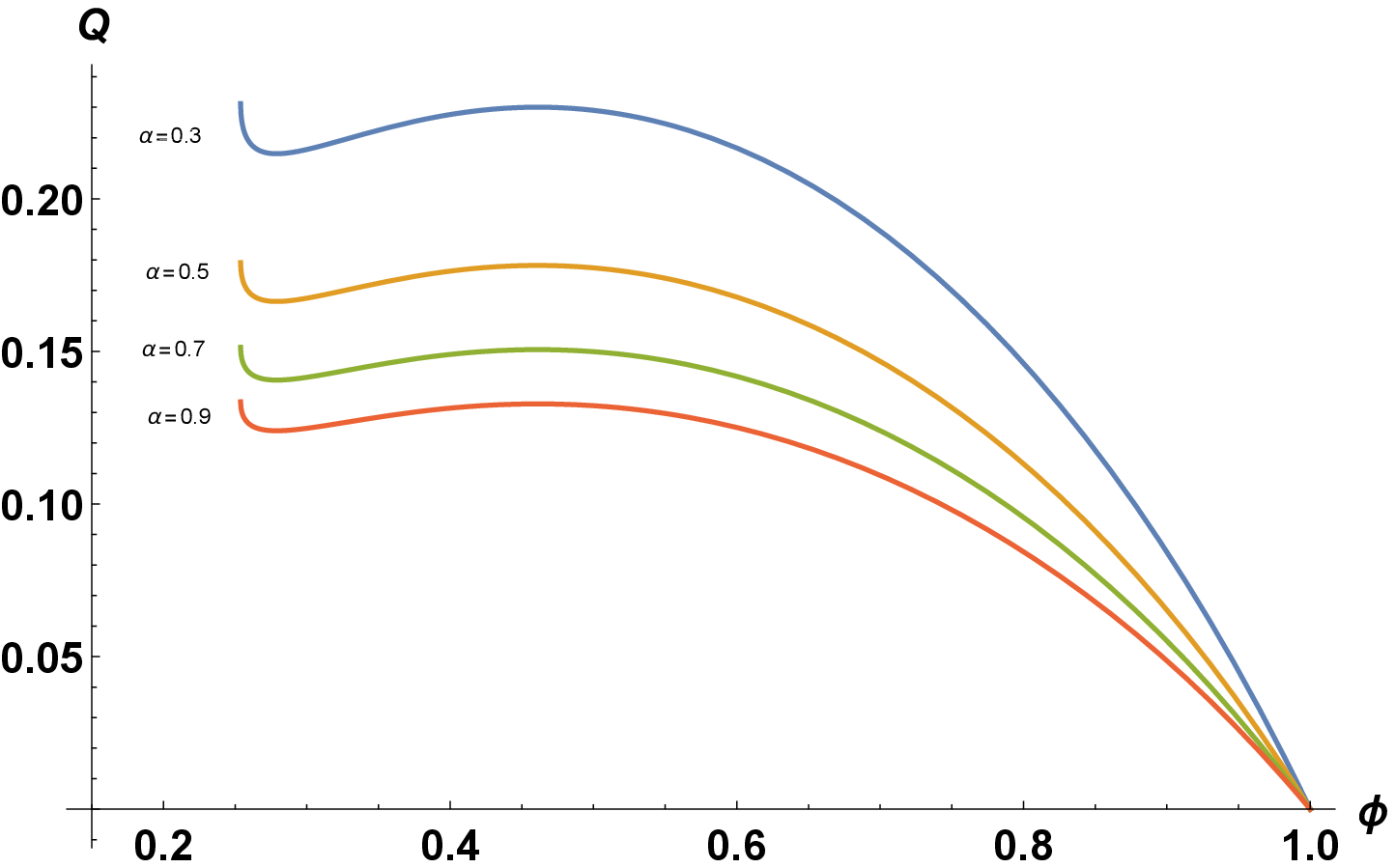}
	\includegraphics[width=8cm]{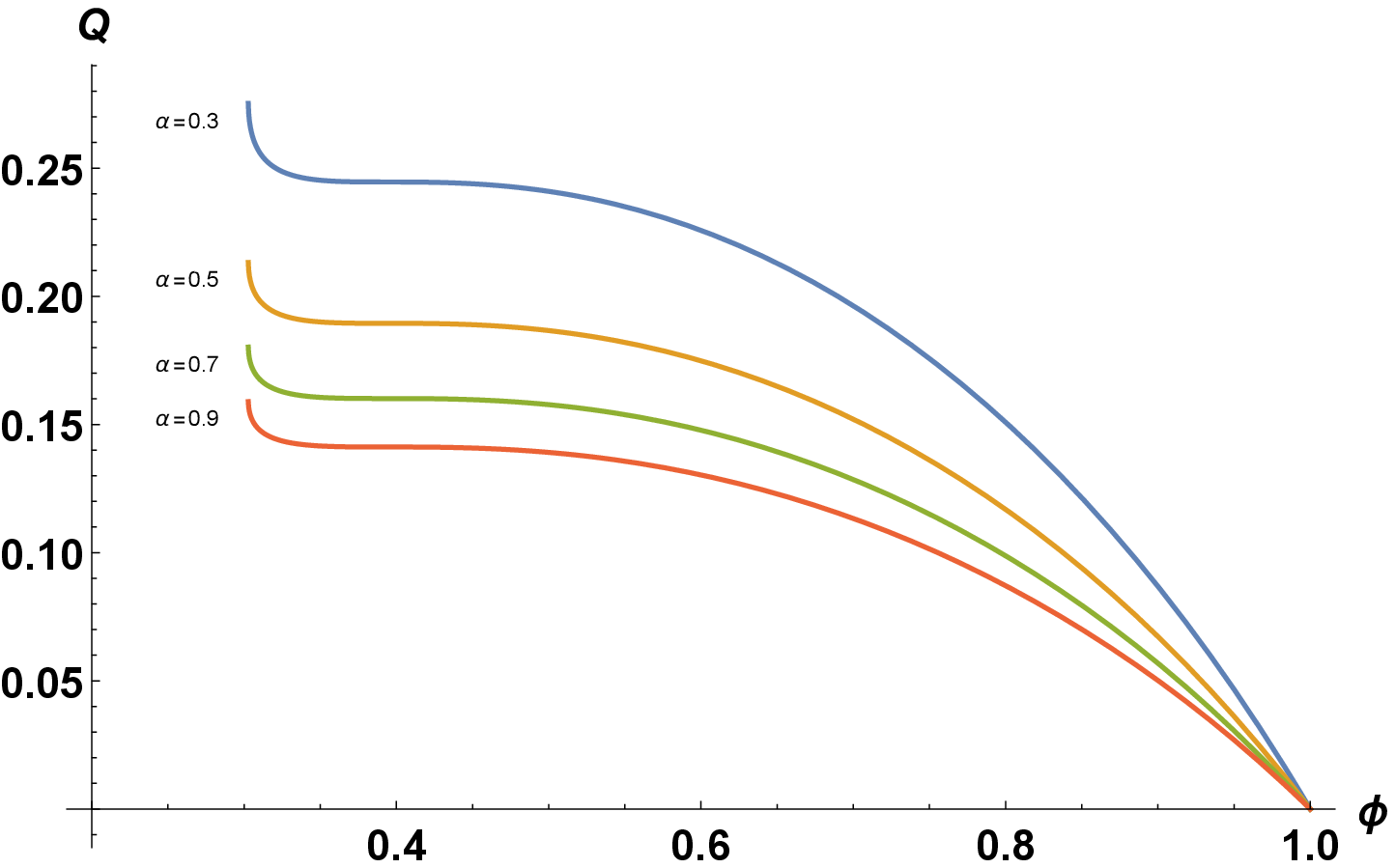}
	\includegraphics[width=8cm]{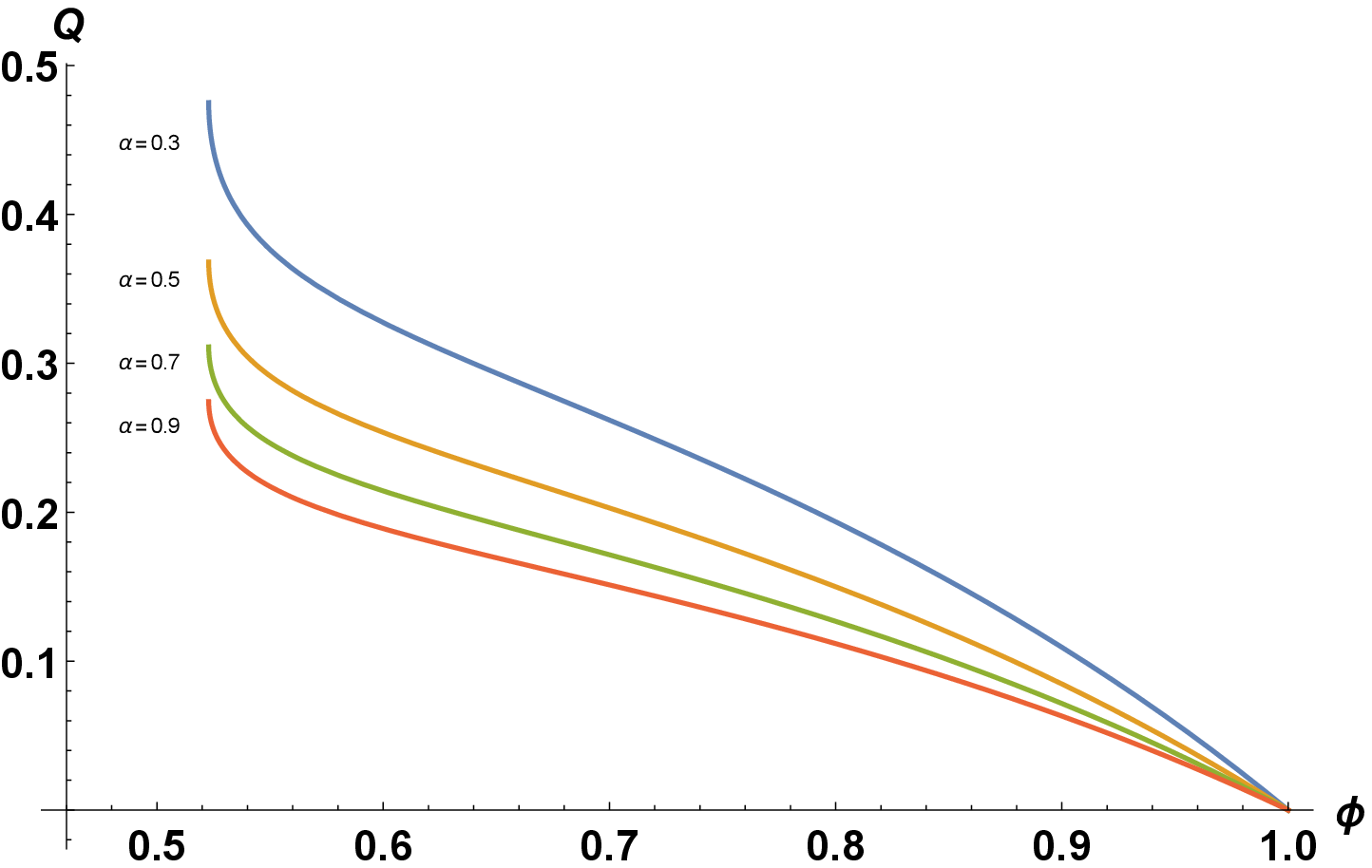}
		\caption{Isothermal curves for different values of the EUP
			parameter. (Top-left) $T>T_{c}$. (Top-right) $T=T_{c}$. (Bottom) $T<T_{c}$.}
		\label{f2}
	\end{figure}

\begin{figure}
	\centering
		\includegraphics[width=13cm]{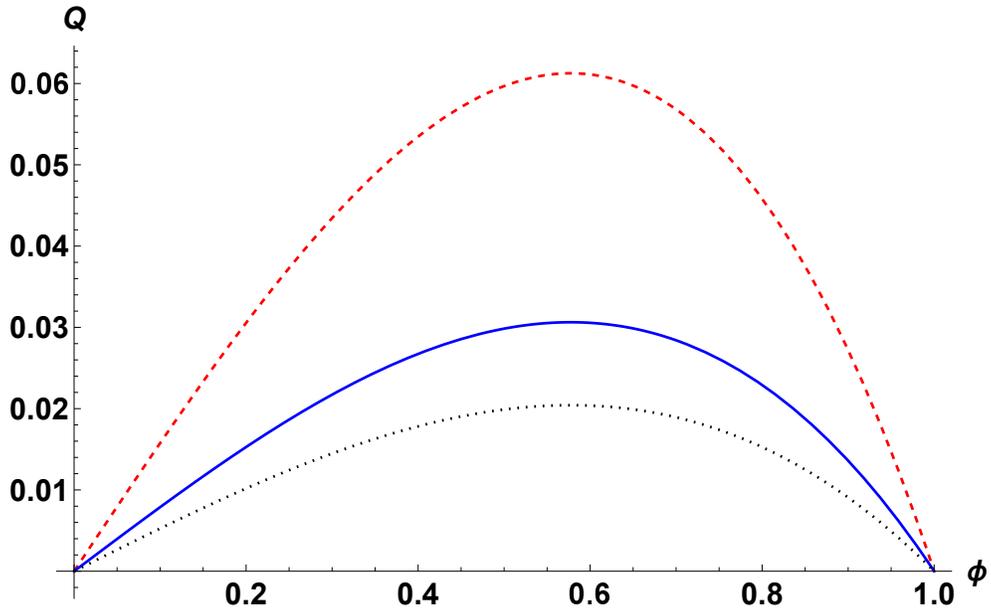}
		\caption{$Q-\Phi $ diagram of RN black hole for $\alpha =0$.
			Red dashed, blue solid and black dotted curves correspond to $T=0.5, 1$ and
			$1.5$, respectively.}
		\label{fb}
	\end{figure}

	In Fig.\ref{f1}, we present the $Q-\Phi $ diagram for the EUP case. It
	is clear that the $Q-\Phi $ diagram resembles to the $P-V$ diagram of vdW
	fluids. The temperatures of curves increase from top to bottom. The upper
	isotherm ($T=0.28<T_{c}$) shows the ideal gas behaviour. The middle isotherm
	corresponds to critical temperature. For $T=0.3>T_{c}$, the lower isotherm
	shows a vdW liquid-gas like phase transition with an unstable region, namely
	oscillating part corresponds to
	$\left (\frac{\partial Q}{\partial \Phi}\right )_{T}>0$. This behaviour
	is similar to AdS black holes and vdW fluids with $(T,P,V)$ as state parameters
	\cite{Kubiznak2012}. In Fig. \ref{f2}, we also investigate effects of the
	parameter $\alpha $ on isothermal curves. It can be seen that
	$\alpha $ does not change the behaviour of the isotherms, but it changes
	the positions of isotherms.\footnote{At this point, we would like to denote
		the differences between $(P,V,T)$ and $(Q,\Phi ,T)$ state parameters. The
		ideal gas and vdW like behaviours of $(Q,\Phi ,T)$ systems occur for
		$T<T_{C}$ and $T>T_{C}$, respectively while the ideal gas and vdW phase
		transition behaviours of $(P,V,T)$ systems occur for $T>T_{C}$ and
		$T<T_{C}$, respectively (please see Figs.~2 and~6 in Ref.~\cite{Kubiznak2012}). In order to investigate the AdS like phase structure
		for EUP corrected case, we consider the analogies
		$P\leftrightarrow Q$, $\Phi \leftrightarrow V$ and
		$T\leftrightarrow \beta =1/T$ where $\beta $ is inverse temperature. In
		fact, we choose inverse temperature $\beta $ as the state parameters rather
		than temperature. In this case, $Q-\Phi $ criticality is more similar to
		conventional $P-V$ criticality, i.e., ideal gas case for
		$\beta >\beta _{C}$ and vdW like phase transition for
		$\beta <\beta _{C}$.} In order to understand the EUP effects, we also plot
	$Q-\Phi $ diagram for $\alpha =0$. Comparing Figs.\ref{f1} and \ref{fb}, it can be seen that the EUP effects dramatically change the behaviour
	of $Q-\Phi $ diagrams. There is not any vdW like phase transition in the
	absence of EUP effects, but it is still worth to investigate the
	$Q-\Phi $ diagrams for $\alpha =0$. In Fig.~\ref{fb}, each isotherm has
	a maximum value $Q_{max}=\frac{1}{6\sqrt{3}\pi T}$ for
	$\Phi =1/\sqrt{3}$. While the isotherms have an unstable part with
	$\left (\frac{\partial Q}{\partial \Phi}\right )_{T}>0$ for
	$\Phi <1/\sqrt{3}$, they have stable part with
	$\left (\frac{\partial Q}{\partial \Phi}\right )_{T}<0$ for
	$\Phi >1/\sqrt{3}$.
	
	%
	%
	\begin{figure}
	\centering
		\includegraphics[width=13cm]{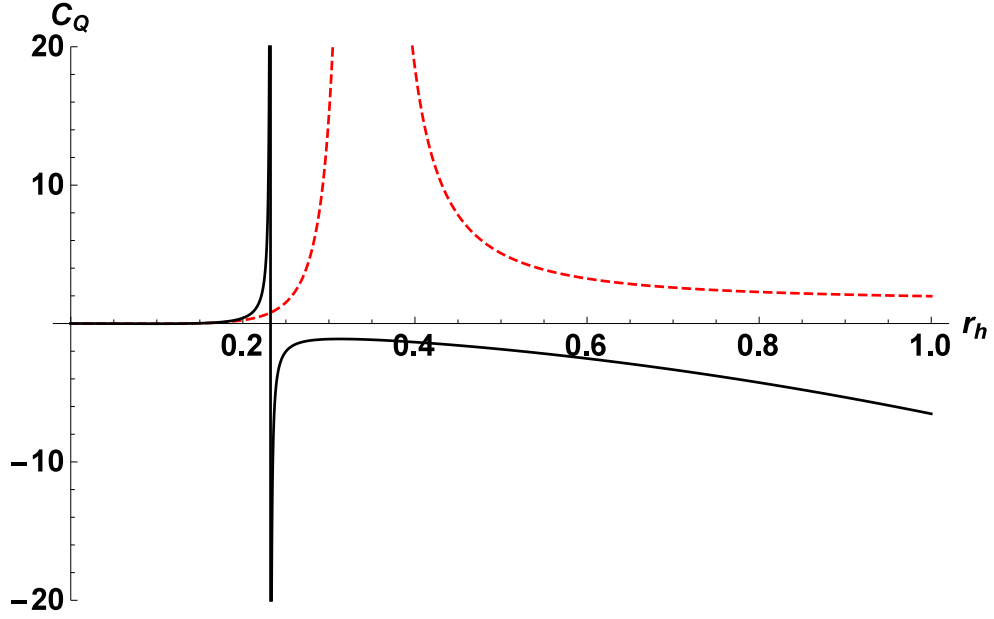}
		\caption{Semiclassical (black solid line) and EUP modified (red
			dashed line) heat capacities versus $r_{h}$. We take $Q=Q_{c}$.}
		\label{f3}
	\end{figure}
	%
	\begin{figure}
	\centering
		\includegraphics[width=13cm]{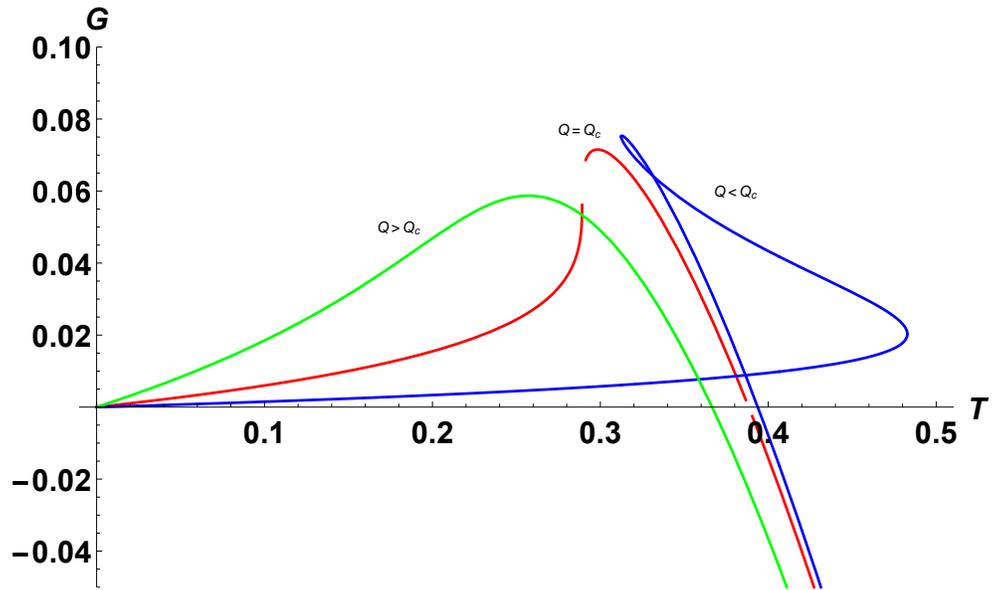}
		\caption{EUP modified Gibbs free energy versus temperature. We
			take $\alpha =1$.}
		\label{f4}
	\end{figure}
	
	\begin{figure}
	\centering
		\includegraphics[width=13cm]{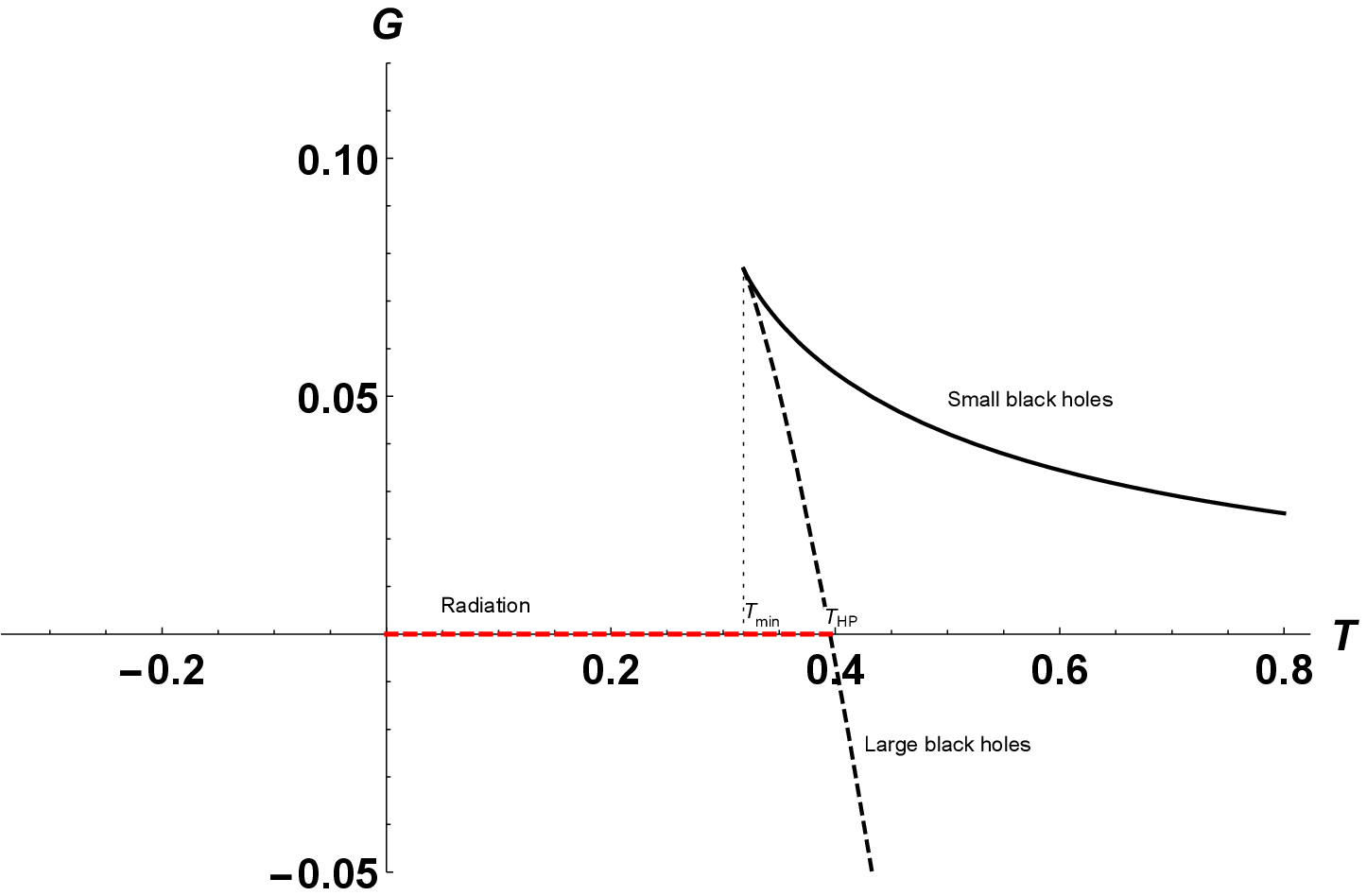}
		\caption{EUP modified Gibbs free energy of Schwarzschild black
			hole versus temperature. We take $\alpha =1$.}
		\label{f5}
	\end{figure}
	
	Moreover, we plot both semiclassical and EUP corrected heat capacities
	in Fig. \ref{f3}. EUP corrected heat capacity has a different behaviour
	for large event horizon values. It diverges near the critical point
	$r_{c}=Q_{c}/\Phi _{c}\sim 0.34$. In contrast to semiclassical case, EUP
	corrected heat capacity is positive for the large values of $r_{h}$ and
	unstable black holes disappear. So EUP corrected heat capacity shares the
	same behaviours with heat capacity of RN-AdS black holes near the critical
	points \cite{Niu2012}.
	
	Finally, we consider the behaviour of Gibbs free energy. In the non-extended
	phase space, Gibbs free energy can be defined by \cite{Ma2016}
	%
	\begin{equation}
		G=M-TS-\Phi Q=
		\frac{\left (Q^{2}-r_{h}^{2}\right )\left [-8\alpha r_{h}^{2}+\left (1+4\alpha r_{h}^{2}\right )\ln \left (1+4\alpha r_{h}^{2}\right )\right ]}{16\alpha r_{h}^{3}}.
		\label{GibbsFree}
	\end{equation}
	In Fig. \ref{f4}, we plot the Gibbs free energy as a function of the temperature
	for the different values of the charge. For $Q<Q_{c}$, the Gibbs free energy
	shows the characteristic swallow tail behaviour which implies the first
	order phase transition.

\section{Hawking-Page like phase transition of Schwarzschild black holes}

\label{sec4}
	
	Another important phase transition is the Hawking-Page phase transition.
	In Ref.~\cite{Hawking1983}, Hawking and Page studied the thermodynamics
	properties of Schwarzschild-AdS black holes. They also showed that the
	canonical ensemble exists for Schwarzschild-AdS black holes, unlike the
	counterparts in flat space. They also found a phase transition between
	black hole and thermal AdS space.
	
	Now, we show a Hawking-Page like phase transition for the EUP case. For
	$Q=0$, EUP corrected Gibbs free energy of Schwarzschild black hole is given
	by
	%
	\begin{equation}
		G=M-TS=\frac{r_{h}}{2}-
		\frac{\left (1+4\alpha r_{h}^{2}\right )\ln \left (1+4\alpha r_{h}^{2}\right )}{16\alpha r_{h}}.
		\label{GibbsFreeEnergySch}
	\end{equation}
	
	Using Eq. (\ref{GibbsFreeEnergySch}), we plot the Gibbs free energy with
	respect to temperature in Fig. \ref{f5}. The figure clearly reveals the
	Hawking-Page like phase transition for the Schwarzschild black hole in flat
	spacetime. Just like AdS case, there are two branches. The upper branch
	(black solid line) corresponds to negative heat capacity and defines the
	small unstable black holes. The lower branch (dashed black line) defines
	the large stable black holes with positive heat capacity. Just like AdS
	case, there is a minimum temperature $T_{min}=\sqrt{\alpha}/\pi $. Below
	this temperature, black holes cannot exist. On the other hand, radiation-black
	hole phase transition occurs at Hawking-Page temperature $T_{HP}$. From
	the $G=0$ condition, we find $T_{HP}=1.24\sqrt{\alpha}/\pi $.

\section{Conclusions}
 
\label{sec5}
	
	In this paper, we focused on the phase transitions of RN and Schwarzschild
	black holes for the EUP case. Taking $T$, $Q$ and $\Phi $ as the state
	parameters, we studied the $Q-\Phi $ criticality of charged black holes.
	We obtained the critical points and showed that there is a vdW like phase
	transition for the charged black hole in the EUP case. We also calculated
	the heat capacity at constant charge and Gibbs free energy. We plotted
	Gibbs free energy and found the characteristic swallow tail behaviour which
	is similar to vdW fluids. Finally, we found Hawking-Page like phase transition
	for the Schwarzschild black hole. In summary, black holes in flat spacetime
	have the same phase structures with the black holes in AdS spacetime in
	the presence of EUP effects.
	
	The black holes in flat spacetime may share similar thermodynamic behaviours
	with black hole in AdS spacetime due to the non-negligible EUP effects
	at large distance scales. Just like cosmological constant $\Lambda $, EUP
	effects play the same role for the thermodynamic stability and the phase
	transition of black holes in flat space. Moreover, Renyi entropy of a black
	hole may be obtained from EUP \cite{MoradpourEUP2019}. On the other hand,
	black holes in flat spacetime has similar thermodynamic behaviours with
	the AdS black holes for the Tsallis-Renyi approach
	\cite{Czinner2016,Promsiri2020}. Therefore, our study also reveals the
	more similarities between the EUP and Renyi entropy.
	
	It is also possible to find similar phase structures in the flat spacetime
	by considering the black holes in a cavity
	\cite{Carlip2003,GuoCavity2013,Wang2021}. The different approaches may
	play the same role at the large distance scales. Therefore, we may consider
	that they are related to each other.
	
\section*{CRediT authorship contribution statement}
      Özgür Ökcü: Conceptualization, Writing - original draft, Writing - review \& editing. Ekrem Aydiner: Supervision, Validation.
	
\section*{Declaration of competing interest}
		The authors declare that they have no known competing financial
		interests or personal relationships that could have appeared to influence
		the work reported in this paper.
		
\section*{Data availability}
		No data was used for the research described in the article.

\section*{Acknowledgments}
The authors thank the anonymous reviewer for his/her helpful and constructive comments. This work was supported by Istanbul University Post-Doctoral Research Project: MAB-2021-38032.

\end{document}